\documentclass[a4paper]{jpconf}
\usepackage{graphicx}
\usepackage[T1]{fontenc}

\usepackage{color}							
\usepackage{float}	
\usepackage{amsmath}
\usepackage{amsfonts}
\usepackage{amssymb}
						
\usepackage{subfig}
\usepackage{a4wide}								
\usepackage{fancyhdr}							
\usepackage{array}
\usepackage{paralist}	
\usepackage{dsfont}
\usepackage{epstopdf}

\usepackage{color}
\usepackage[font={footnotesize,it}]{caption}

\numberwithin{equation}{section}

\begin{document}
\title{Spreading dynamics on networks: the role of burstiness, topology and non-stationarity}

\author{D\'{a}vid X. Horv\'{a}th$^1$ and J\'{a}nos Kert\'{e}sz$^{1,2,3}$}

\address{$^1$ Institute of Physics, BME, Budapest,
Budafoki \'{u}t 8, H-1111, Hungary}
\address{$^2$ Center for Network Science, CEU, Budapest, N\'{a}dor utca 9, H-1051, Hungary}
\address{$^3$ Department of Biomedical Engineering and Computational Science, Aalto University}
\ead{janos.kertesz@gmail.com}

\begin{abstract}
Spreading on networks is influenced by a number of factors including different parts of the inter-event time distribution (IETD), the topology of the network and non-stationarity. In order to understand the role of these factors we study the SI model on temporal networks with different aggregated topologies and different IETD-s. Based on analytic calculations and numerical simulations, we show that if the stationary bursty process is governed by power-law IETD, the spreading can be slowed down or accelerated as compared to a Poisson process; the speed is determined by the short time behaviour, which in our model is controlled by the exponent. We demonstrate that finite, so called "locally tree-like" networks, like the Barab\'{a}si-Albert networks behave very differently from real tree graphs if the IETD is strongly fat-tailed, as the lack or presence of rare alternative paths modifies the spreading. A further important result is that the non-stationarity of the dynamics has a significant effect on the spreading speed for strongly fat-tailed power-law IETD-s, thus bursty processes characterized by small power-law exponents can cause slow spreading in the stationary state but also very rapid spreading heavily depending on the age of the processes.
\end{abstract}

\section{Introduction}
Processes on complex networks may consist of events on links with a specific starting time and duration. When no event happens, the link can be considered as temporarily absent. Typical examples are communication networks or airline connections. Phenomena related to such processes are best described within the framework of temporal networks~\cite{Holme-Saramaki}.

Many important time dependent processes taking place on various complex networks exhibit bursty nature, i.e., the events of such processes are distributed very heterogeneously in time. Besides the burstiness of natural phenomena like earthquakes or the firing of neurons, processes in several man-made or sociological systems like contact patterns in human communication networks are known to be bursty as well. One of the most important characteristics of this bursty dynamics is the inter-event time distribution (IETD), which has a power-law form in a broad regime of time in many systems of interest \cite{barabasinature}.

In recent years, it has been a major challenge to understand how burstiness influences spreading phenomena on networks. Despite of the considerable effort devoted to this question the answers are still contradictory.
Using empirical contact sequences to study random walks, Starnini et al.  \cite{starnini} showed that burstiness present in empirical data slows down considerably the random walk exploration. By analyzing the spread of e-mail worms using data of a commercial provider \cite{slowdown2} and by the investigation of rumor spreading in a mobile communication network  \cite{smallbutslowworld} the burstiness was found to slow down spreading, whereas in the work of Rocha et al. \cite{Rocha_fast} similar studies using human network of sexual contacts suggest that bursty temporality accelerates disease spread. While analytically solvable models provide some insight into how burstiness influences spreading, they are usually not flexible enough to incorporate important details like network topology, however,  according to the analytically solvable SI model of Jo et al. \cite{egzakt}, the power-law IETD steered dynamics always results in faster spreading than the corresponding Poissonian-like dynamics that also points towards fast bursty spreading. Some papers present a more complex picture. Based on analysis of mobile phone calls, Miritello et al. \cite{Miritello} claim that long waiting times hinders but group conversations favour the spread of information and the overall spreading is determined by the competition of these two factors. In the paper of Holme et al. \cite{HL} the role of the beginning and end time of communication sequences is pointed out and in the work of Rocha et al. \cite{SIR} non-stationary bursty processes yield faster than Poissonian spreading in SI and SIR models in a large regime of parameters.

In this paper we wish to contribute to better understanding the influencing factors of the spreading on networks by studying the so called Susceptible-Infected (SI) model~\cite{Barrat} with bursty temporality and some particular topological properties affecting spreading. We will also point out the effect of the age of the process, i.e., that of non-stationarity. While these models are still simplifications of the complex, real world situations with many correlations ignored here, we hope that 
by systematic investigations we can identify and understand the role of each factor in the outcome of the spreading and in its speed.
The paper is organized as follows: In the next section we define the model and the quantities of interest. Then we discuss the case of the Cayley tree, where some analytical results can be achieved. In \emph{Section 4} we compare the spreading behavior of Barab\'{a}si-Albert trees and networks in the stationary state. The following section contains the results on the role of non-stationarity. The paper ends with a Summary and Discussion.

\section{Model definition and quantities of interest}

\subsection{Temporal and aggregated networks}

Processes are defined on temporal networks, in which the network topology enters the model through the aggregated network. The aggregated network is defined as the static graph consisting of the nodes of the temporal network and the links where at least one event takes place. For aggregated networks, we consider Cayley trees (regular trees), where analytical calculations are possible. Cayley trees are trees, i.e., networks without loops in which all the nodes have exactly the same number of neighbors or equivalently the same degree denoted by $k$. We consider also Barab\'{a}si-Albert (BA) networks \cite{BAcikk}, which share some properties of real world networks. In the BA model the network is grown by adding nodes one by one with linear preferential attachment \cite{PrefAtt}.
BA networks are known to behave locally as trees as they have very low number of loops \cite{loops}. By changing the degree $m$ of the new born nodes in the BA network, we can compare spreading on a tree ($m=1$) and on graphs with loops ($m >1$), however keeping the locally tree-like feature. 

In our model the inter-event times are the time intervals between the events between a given pair of nodes in the temporal network. As already mentioned, the inter-event times are characterized by the IETD, which are identically distributed, independent (iid) random variables for all edges in our model, thus we consider on the links independent renewal processes with iid inter-event times. In general, the process is non-Markovian and it takes (infinitely) long time to reach stationarity. If the IETD is exponential, we have a Poisson process, which, due to its "memoryless" character, is instantaneously stationary.

\subsection{Time distributions}

To take the burstiness of the temporal processes into account, we use the Pareto distribution for the inter-event time distribution:
\begin{equation} \label{paretodist}
p_{pow}(t)=
\begin{cases}
t_{min}^{\alpha}\alpha\frac{1}{t^{\alpha+1}}, & \text{if}\ t \geq t_{min}\\
0, & \text{otherwise}.
\end{cases}
\end{equation}
In Eqn. \eqref{paretodist} $t_{min}$ denotes the lower cut-off and the $\alpha$ exponent determines the the fatness of the tail of the distribution. To demonstrate the impact of burstiness, it is worth comparing the effect of power-law and exponential IETD on the spreading. For this reason we prescribe the same mean for the applied IETD-s denoted by $\tau$ and use modified exponential functions with lower cut-off too:
\begin{equation}\label{expdist}
p_{exp}(t)=
\begin{cases}
\frac{\alpha}{\tau} e^{-\alpha \frac{t-t_{min}}{\tau}}, & \text{if}\ t \geq t_{min}\\
0, & \text{otherwise}.
\end{cases}
\end{equation}
This choice enables to investigate the importance of the lower cut-off \cite{egzakt}, \cite {oktobericikk}.
Since the mean is kept fixed in these distributions, the $\alpha$  parameters can be expressed by means of $\tau$ and $t_{min}$ with their ratio $v=\frac{\tau}{t_{min}}$. Then one gets $\alpha= \frac{v}{v-1}$ for both the Pareto and the modified exponential case. For the exponential distributions an important limiting case is when $\alpha=1$ or $t_{min}=0$, which yields a homogeneous Poisson process.

The dynamical process on a link is characterized by the IETD, however, the time to be waited by a newly infected node to infect one of its neighbors, which is referred to as the waiting time,  is characterized by the waiting time distribution (WTD), which is the distribution of the time to be waited between a random time (the infection of a node) and an event (transmission). In general, the WTD depends also on the age $T$ of the process \cite{Feller} unless the process is Poissonian, when the IETD and WTD coincide. For the stationary case ($T \to \infty$), the WTD $p'(t)$ from a given IETD p(t) with mean $\tau$  is given as \cite{Feller}: 

\begin{equation}\label{wtd}
p'(t)=\frac{1}{\tau}\int_t^{\infty} p(t') \,dt' .
\end{equation}
In the literature on spreading stationarity has been usually implicitly assumed by using \eqref{wtd} for the WTD (\cite{egzakt}, \cite{Miritello}, \cite{oktobericikk}, \cite{iribarren}), however,  the problem of stationarity will prove to be of fundamental importance regarding the spreading speed.

\subsection{SI model and averages}

To model bursty spreading we choose the simplest Susceptible-Infected (SI) model, in which the links between the nodes are present or active at specific instants of time corresponding the temporal network picture with no duration of the events.  In the SI spreading model, at t=0 we infect a node and from this instant the disease propagates with probability 1 if there is an existing link between an infected and a susceptible node.

The basic quantity that characterises the spreading is the curve given by the fraction of infected nodes versus time or in infinite networks the number of infected nodes versus time. This curve can be calculated as the mean of random variables, however, the averaging can be done either by fixing time $t$ and taking the average number or fraction of nodes at this instant, or by fixing the number or fraction of infected nodes and calculating the average elapsed time until the system reaches this given level of infection.
Both ways of averaging provide important and relevant quantities; averaging according to the first method tells us the average level of infection after time $t$, and the second method telling the average time until a given level of infection is reached can be important, e.g. for studies of vaccination or prevention strategies. In the following we will refer to the first method as $\bar{N}_t$ average, and to the second method as $N_{\bar{t}}$ average. These averages are related to each other (see \emph{Appendix A}).

\section{Stationary dynamics on Cayley trees}
As long as the aggregated networks are infinite Cayley trees it is possible to obtain either the whole spreading curve or the early time and late time asymptotics of the spreading curves in the stationary state, therefore we begin with presenting some analytic calculations.
Concerning the $N_{\bar{t}}$ curve we restrict ourselves only to the Poissonian case, whereas the derivation of the $\bar{N}_t$ curve is of more general nature. 
For the derivation of both curves we consider Cayley trees with $k$ branches per nodes or nodes with degree $k$ and $N$ denotes the number of infected nodes.

To obtain the $N_{\bar{t}}$ curve for the Poissonian dynamics, one needs to consider the number of \emph{active} links $N_{l, a}(i)$, i.e., the number of links between infected and susceptible nodes after the \emph{i}-th node is infected. 
It is easy to write down the recursion $N_{l, a}(i+1) =N_{l, a}(i)+k-2$, which has the solution $N_{l, a}(i)= ik-2(i-1)$. 
Since the Poisson process is memoryless, the average value of the next infection time given that there are $n$ active links is merely the expected value of the minimum of $n$ iid exponential random variables with mean $\tau$, which is $\frac{\tau}{n}$, therefore the following equation can be written for the average time of the $i$-th infection:

\begin{equation} \label{Cayleyrecursion}
\bar{t}(i)=\sum_{j=1}^{i} \frac{\tau}{jk-2(j-1)},
\end{equation}
whose solution is  
\begin{equation} \label{Cayleyrecursionmo}
\bar{t}(i)=\tau \frac{\psi^{(0)}(\frac{i(k-2)+k}{k-2}) -\psi^{(0)}(\frac{k}{k-2})}{k-2},
\end{equation}
where $\psi^{(0)}(x)=\frac{d}{dx} ln(\Gamma(x))$.
 As $\frac{d}{dx} ln(\Gamma(x)) \approx ln(x)$ for large $x$ one can see that the $N_{\bar{t}}$ inverse curve is of the form $N(\bar{t}) \approx e^{\frac{k-2}{\tau}\bar{t}}$ for large $t$. As the Poissonian assumption was heavily used in this derivation, this reasoning cannot be generalized to other IETD-s.

For the $\bar{N}_t$ curves the theory of the Bellman-Harris branching processes \cite{iribarren}, \cite{Harris} can be made use of. For this, let $N_l$ denote the number of \emph{infected} links, that is the number of links in the aggregated network having at least one infected node on its ends and let $N_l^*$ be the number of such links in one branch of the initially infected node. It is easy to see, that the number of infected nodes $N$ is expressed as:
\begin{equation} \label{node-pseudonode0}
N=\frac{N_l-k}{k-1}+1.
\end{equation}
Let then $N_l^*(t)$ be a stochastic process, so for a given $t$,  $N_l^*(t)$ is a random variable telling the number of \emph{infected} links in one branch for a given time.  Supposing first that each node infects its neighbors after a fixed value of $w$, we can write the following equation:

\begin{equation}\label{w}
N_l^*(t)=
\begin{cases}
1, & \text{if}\ t \leq w\\
1+\sum_{i=1}^{k-1} N_{l, i}^{*} (t-w),                                                     & \text{if}\ t > w. \\
\end{cases}
\end{equation}
Here $N_{l, i}^{*}(t)$ denotes identical copies of $N_{l}^*(t)$ . Using the generating function of $N_{l}^*(t)$ defined as $F(t,z)=\sum_{n=0}^{\infty} \mathds{P} \left[ N_{l}^*(t)=n \right] z^n$ Eqn. \eqref{w} can be rewritten as
\begin{equation}\label{genfvw}
F(t,z)=
\begin{cases}
z, & \text{if}\ t \leq w\\
z+F(t-w,z)^{k-1},                                                     & \text{if}\ t > w. \\
\end{cases}
\end{equation}
Integrating Eqn. \eqref{genfvw} over $w$ with its weight function, which is the WTD, and differentiating $F(t,z)$ with respect to $z$ and finally setting $z=1$ we arrive for $\bar{N}_l^*(t)$ denoting the average of  $N_{l}^*(t)$ at

\begin{equation}\label{Ncsillagint}
\bar{N}_l^*(t)=1+\int_0^t (k-1)\bar{N}_l^*(t-w) p'(w) \,dw .             
\end{equation}
It is worth taking the Laplace transform of Eqn. \eqref{Ncsillagint} and after some arrangement one obtains 
\begin{equation}\label{Laplacealaposszefugges}
\tilde{N}_l^*(s) =\frac{1}{s} \frac{1}{1-(k-1)\tilde{p}'(s)},
\end{equation}
where $\tilde{N}_l^*(s)$ and $\tilde{p}'(s)$ denote the Laplace transform of $\bar{N}_l^*(t)$ and $p'(t)$.
As $\bar{N_l}=k \bar{N}_l^*$ and from \eqref{node-pseudonode0}, we get for the average number of infected nodes the following relationship:
\begin{equation} \label{node-pseudonode}
\bar{N}(t)=\frac{k \bar{N}_l^*(t)-k}{k-1}+1,
\end{equation}
where the linearity of \eqref{node-pseudonode0} is made use of.

If the dynamics is Poissonian, $\tilde{p}'(s)= \frac{1}{1+\tau s}$ and the inverse transformation of Eqn. \eqref{Laplacealaposszefugges} is simple, yielding
\begin{equation}\label{LaplacePoiN}
\bar{N}(t)=\frac{k}{k-2}e^{\frac{k-2}{\tau}t}-\frac{2}{k-2}.
\end{equation}
Comparing this formula with the $N_{\bar{t}}$ average (Eqn. \eqref{Cayleyrecursionmo}), we can conclude that for large $t$ both averages predict exponential growth according to $e^{\frac{k-2}{\tau}t}$.

If, however, the IETD is a power-law or an exponential with a lower cut-off,  only the asymptotic behaviour can be calculated analytically. 
The Laplace transform of the WTD derived from the Pareto distribution and the transform of WTD of the modified exponential IETD  are as follows:
\begin{equation}\label{powlaplace}
\tilde{p}'_{pow}(s)=\frac{1}{\tau s} \left(1-e^{-t_{min} s} \right)+ \frac{t_{min}}{\tau} \varphi_{-\alpha}(t_{min} s),
\end{equation}
\begin{equation} \label{explevaglaplace}
\tilde{p}'_{exp}(s)=\frac{1}{\tau s} \left( 1-e^{-t_{min} s} \right)+\frac{e^{-t_{min}s}}{\alpha+\tau s},
\end{equation}
where 
\begin{equation}\label{Misra}
\varphi_{-\alpha}(z)=\int_1^{\infty} x^{-\alpha} e^{-xz} \,dx,
\end{equation}
which can be expressed with the incomplete gamma function:
\begin{equation}\label{Misragamma}
\varphi_{-\alpha}(z)=z^{\alpha-1}\Gamma(1-\alpha, z).
\end{equation}

The early time spreading dynamics ($t << t_{min}$) can be obtained by inverse transforming the asymptotic form of $\tilde{N}^*(s)$ as $s \rightarrow \infty$.
Due to the asymptotics of the incomplete gamma function \cite{Abramowitz} for large z it yields:
\begin{equation} \label{ordo}
 \varphi_{-\alpha}(z)=\frac{e^{-z}}{z}\left( 1+\frac{\Gamma(1-\alpha)}{\Gamma(1-\alpha-1)}\frac{1}{z}+ \frac{\Gamma(1-\alpha)}{\Gamma(1-\alpha-2)}\frac{1}{z^2}+ \mathcal{O}(\frac{1}{z^3})\right).
\end{equation}
It is easy to see that the asymptotics of both $p'_{pow}(s)$ and $p'_{exp}(s)$ is determined by the first term in Eqn. \eqref{powlaplace} and \eqref{explevaglaplace}. The emergence of this term in the Laplace transforms comes from the lower cut-off in the IETD, which results in a constant term in the WTD. 
As a consequence,  in case of either the Pareto or the modified exponential IETD, the Laplace transform of the WTD has the same asymptotics for large $s$, resulting in practically the same curve for sufficiently small times:
\begin{equation}\label{Cfakisido}
\tilde{N}(t)=\frac{k}{k-1}  e^{\frac{k-1}{\tau}t}  - \frac{1}{k-1},  \text{for $t  << t_{min}$}.
\end{equation}
According to Eqn. \eqref{Cfakisido} as long as the $\bar{N}_t$ average is considered and the dynamics is stationary the modified exponential and power-law governed spreading is always faster on average than the simple Poissonian spreading if $t$ is sufficiently small and despite of the absence of the $\alpha$ dependence in the asymptotic form of $N(t)$ the size of  lower cut-off determines the domain of validity of Eqn. \eqref{Cfakisido} as $s t_{min} >>1$ has to hold when approximating $p'(s)$.

\begin{figure}[H]
\centering
\includegraphics[width=8.5 cm]{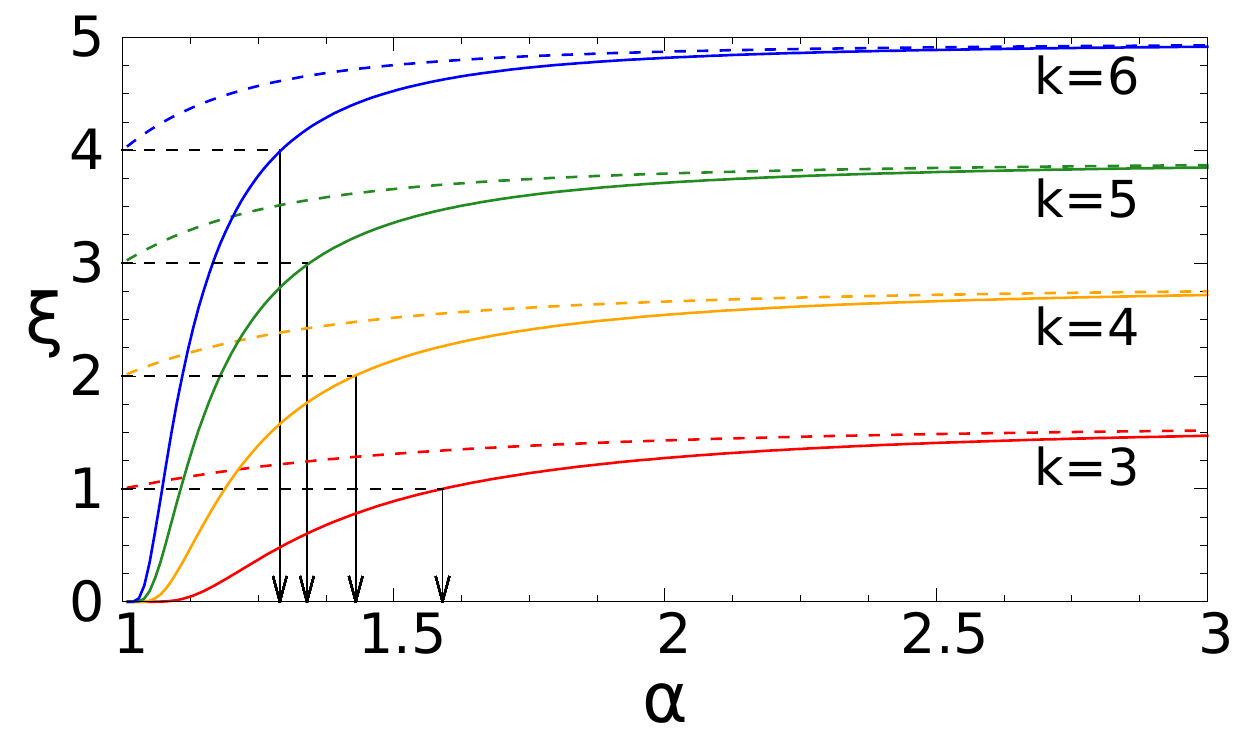} 
\caption{\textit{Time constant $\xi$ of the exponential growth for late time spreading dynamics ($\tau=1$) for different values of $\alpha$, where $\alpha$ is expressed as $\frac{v}{v-1}$, $v$ being $\frac{\tau}{t_{min}}$. The colored continuous and dashed lines correspond to the power-law and modified exponential case, and different colors to different Cayley trees with certain number of neighbors $k$: red: $k=3$, orange: $k=4$, green: $k=5$, blue: $k=6$. The x coordinate of the intersection of the horizontal dashed lines and the continuous colored lines corresponds to $\alpha_C$ at which the slower than Poissonian-faster than Poissonian transition happens for power-law governed dynamics. The values of $\alpha_C$ are 1.5900; 1.4276; 1.3454 and 1.2924 for $k=$ 3,4,5 and 6.}}
\label{Cfanagyido}
\end{figure}

The late time asymptotics of $\bar{N}(t)$ is supposed to be also exponential of the form $\bar{N}(t) \approx C e^{\xi t}$ and to calculate $\xi$ the final value theorem \cite{fvt}, \cite{egzakt}  is used, according to which 
\begin{equation}\label{fvteq}
C=\lim_{t \to \infty}  \bar{N}(t) e^{-\xi t}= \lim_{s \to 0} s \tilde{N}(s+\xi).
\end{equation}
To get a meaningful result, the largest real pole of Eqn. \eqref{Laplacealaposszefugges} has to be at the origin by shifting it with $\xi$, thus  $\xi$ is merely the zero of the following equation:
\begin{equation} \label{zero}
1-(k-1)\tilde{p}'(s).
\end{equation}
The zero of this equation, which is unique,  can be calculated only numerically and Fig. \ref{Cfanagyido} displays its value for different $k$ and $ \alpha$ parameters.
From Fig. \ref{Cfanagyido} some important conclusions can be drawn. We recall that $\alpha$ controls the lower cut-off, $t_{min}$ in Eqn.  \eqref{paretodist} and \eqref{expdist}, and that $\alpha \to 1$ means for the exponential case, $t_{min}=0$, i.e., the Poisson process. For the power-law case, $\alpha$ is the power-law exponent, as according to \eqref{paretodist}, $p_{pow}(t) \propto t^{-(\alpha+1)}$ for $t > t_{min}$. For a given $\alpha$ the spreading speed of the process with Pareto IETD is always smaller than the corresponding process with exponential IETD.  The curves in Fig. \ref{Cfanagyido} are monotonously increasing functions of $\alpha$, and larger $\alpha$ yields more compact IETD-s, one can even easily find out from \eqref{paretodist} and  \eqref{expdist} that the $\alpha \to \infty$ limit corresponds to an IETD of the form $\delta(t-\tau)$ and a WTD uniform on $[0, \tau]$. We can then conclude that the more narrow the IETD is, the faster resulted spreading is seen. This result is similar to that presented in \cite{oktobericikk} in which diffusion processes are studied and essentially the second moment of the WTD characterizes the broadness of the distribution. The last important aspect to consider is the spreading speed of the Pareto case compared with the Poisson process. According to Fig. \ref{Cfanagyido} for each $k$ two domains of the $\alpha$ parameter can be distinguished resulting in either faster or slower than Poissonian spreading: For $\alpha \in (1, \alpha_C)$ the power-law spreading is slower than the Poissonian case while for $\alpha \in ( \alpha_C, \infty)$ the power-law spreading is faster. The $\bar{N}_t$ average is not very sensitive to the tail of the WTD. The natural explanation for the above observation is based on the comparison of the short time behaviour of the WTD-s. For power-law WTD there is a constant term with height $1/\tau$ in the interval $[0, t_{min}]$, whereas the exponential WTD is $\frac{1}{\tau}e^{-\frac{t}{\tau}}$. Thus, for large $\alpha$ or equivalently for larger $t_{min}$ the WTD derived from the Pareto IETD has larger probability density for short times, than the exponential distribution resulting in faster spreading. On the contrary, for small $\alpha$ (small $t_{min}$) the spreading is slower for power-law WTD than for the Poisson process.

\newpage
\section{Stationary dynamics on BA networks and trees}

After the investigation of the Cayley tree we turn to the numerical study of spreading on BA networks. In the BA model the network is grown by adding nodes one by one with linear preferential attachment \cite{PrefAtt}. The process is controlled by the number $m$ of links introduced with each node (<k>=2m).  
When simulating SI dynamics on temporal networks with stationary processes, we can apply several simplifications in the algorithm. When a node $A$ gets infected at time $t$, we draw independent random variables ($\xi_1, \xi_2,...\xi_n$) from the WTD and assign the $t+\xi_i$ values, which are possible infection times, to the susceptible neighbors of node $A$ in the aggregated network. Obviously, several neighbors can try to infect a given node but from these attempts only the one with the smallest infection time matters. After assigning contact times to the susceptible neighbors of node $A$  and checking whether they are the smallest ones for a given neighbor, we need to find the node with the earliest next infection time, which is $t'>t$. Thus, with this iterative algorithm we are allowed to deal with events between only infected and susceptible nodes and by assuming stationarity, we can use only the WTD, although the process is defined by the IETD.

\begin{figure}[H]
\centering
\includegraphics[width=7.5cm]{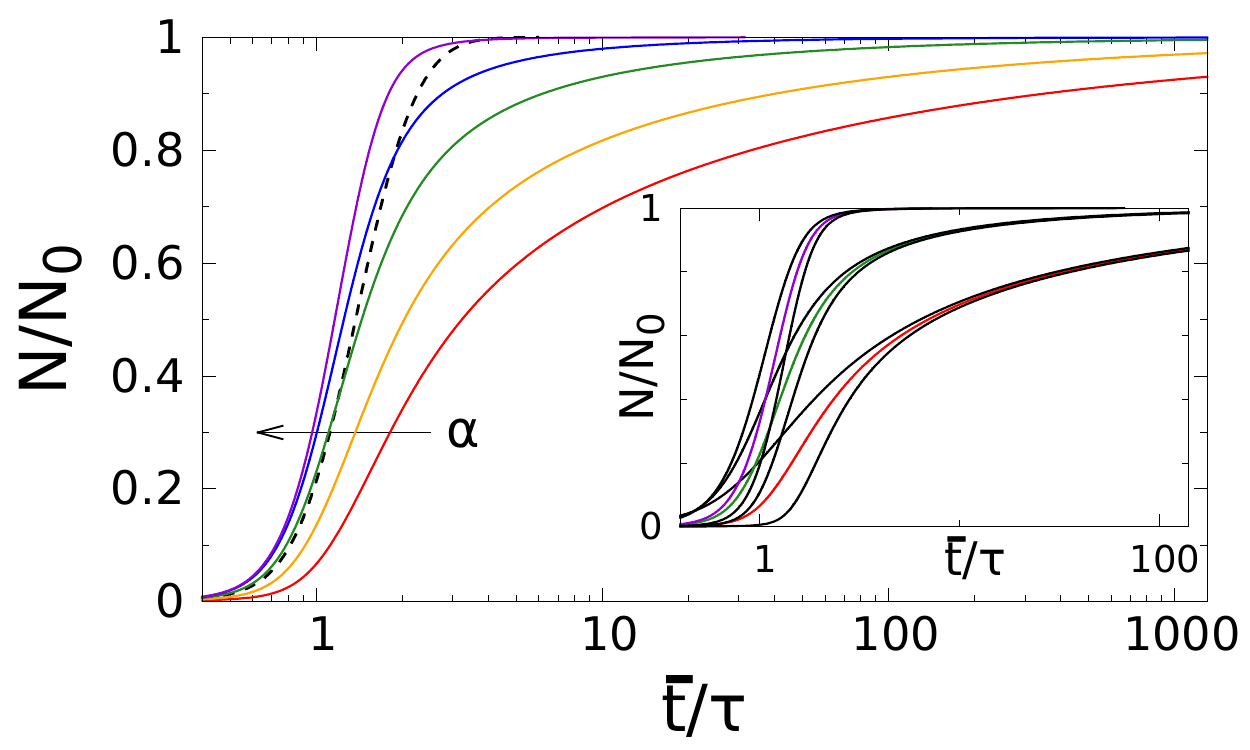}
\caption{\textit{ Fraction of infected nodes $N/N_0$ vs average time $\bar{t}$ (measured in units of mean inter-event time $\tau$) spreading curves of pure exponential  (black, dashed line)  and Pareto (continuous, colored lines)  IETD governed stationary dynamics. Simulation results on BA networks of size $N_0=10^4$ nodes and with average degree $\left<k\right>=4$. The initially infected node was chosen at random in all runs in a way that its degree $k_0$ was larger than 25.  Averages of $2.5 \cdot10^3$ runs in each case and for every fifth run a new network was generated. The colors red, orange, green, blue and violet correspond to $\alpha=1.111, 1.154, 1.25, 1.4 \text{ and } 2$ and $t_{min}=0.1, 0.133, 0.2, 0.286 \text{ and } 0.5$  (in units of $\tau$)  lower cut-offs respectively. The arrow shows increasing values of $\alpha$ with the exception of dashed curve (Poisson process), for which $\alpha=1$. The inset indicates the fluctuations of the curves associated with $\alpha=1.111, 1.25 \text{ and } 2$ with continuous black lines.}}
\label{powwtossz}
\end{figure}

\begin{figure}[H]
\centering
\includegraphics[width=7.5cm]{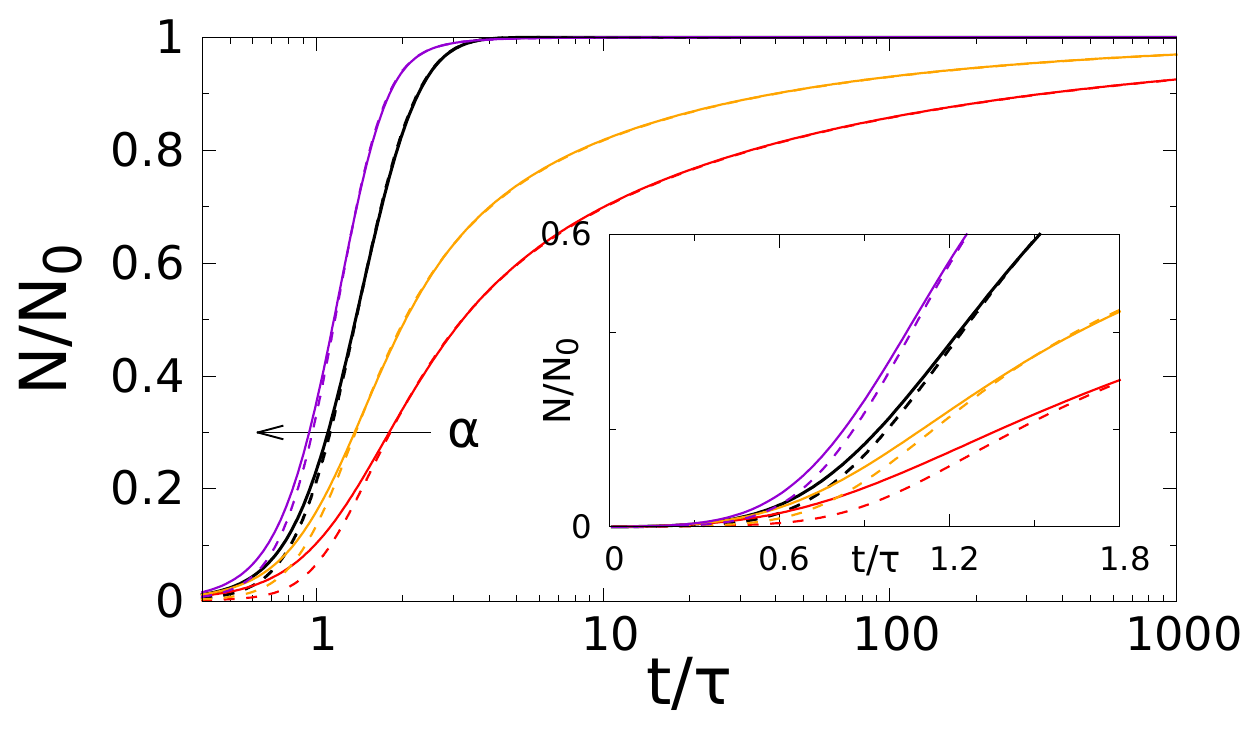}
\caption{\textit{Average fraction of infected nodes $\bar{N}/N_0$ vs time $t$ (measured in units of mean inter-event time $\tau$) spreading curves (continuous lines)  and fraction of infected nodes $N/N_0$ vs average time $\bar{t}$ (again measured in units of $\tau$) spreading curves (dashed lines) of IETD governed stationary Poissonian (black lines) and Pareto (colored lines) dynamics. Simulation results on BA networks of size $N_0=10^4$ nodes with average degree $\left<k\right>=4$. The initially infected node was chosen at random in all runs in a way that its degree $k_0$ was larger than 25.  Averages of $2.5 \cdot 10^3$ runs in each case and for every fifth run a new network was generated.  The colors red, orange and violet correspond to $\alpha=1.111, 1.154 \text{ and } 2$ and $t_{min}=0.1, 0.133 \text{ and } 0.5$  (in units of $\tau$)  lower cut-offs respectively for both $\bar{N}_t$ and $N_{\bar{t}}$ averages. The arrow show increasing values of $\alpha$ with the exception of the dashed-continuous, black curves (second pair from the left, Poisson process), for which $\alpha=1$. The inset indicates the early-time behaviour.}}
\label{powwtossz2}
\end{figure}

In Fig. \ref{powwtossz}, $N_{\bar{t}}$ averages of simulation runs are presented.
In Fig. \ref{powwtossz} all the curves have a saturating part in the asymptotic regime of late times, which is the result of the finite size of the the networks used in the simulations and which is a typical property of the SI model. In finite systems the early and intermediate time dynamics are to be compared with the early and late time dynamics in the infinite system. Focusing on the intermediate regime (intermediate level of infection) in Fig. \ref{powwtossz}, for sufficiently large $\alpha$ rapid, and for sufficiently small $\alpha$ slow spreading is seen for Pareto IETD, and for decreasing $\alpha$ the slowing down is more pronounced. This is obviously a similar behaviour to what we discussed in connection with the Cayley tree and such a similarity also exists with the modified exponential case (here not shown).
The conclusions above are based on $N_{\bar{t}}$ averages, however, the $N_{\bar{t}}$ and $\bar{N}_t$ averages approach each other quickly if the initially infected nodes have high degree, as seen in Fig. \ref{powwtossz2}.  The $N_{\bar{t}}$ curves approach the $\bar{N}_t$ curves always from below, which is understandable considering the fact, that the $N_{\bar{t}}$ averages are influenced by the fat-tailed IETD-s.

\begin{figure}[H]
\centering
\includegraphics[width=7 cm]{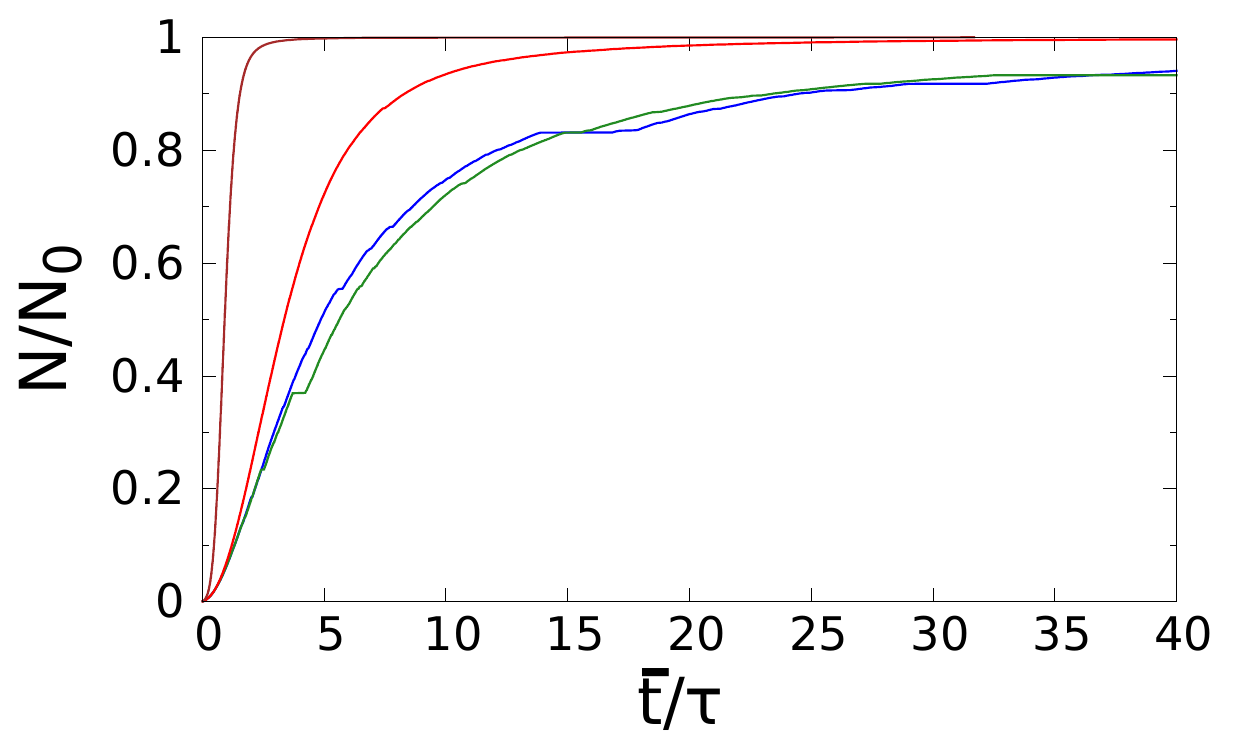} 
\tiny
\caption{\textit{Fraction of infected nodes $N/N_0$ vs average time $\bar{t}$  (measured in units of mean inter-event time $\tau$) spreading curves of power-law governed stationary dynamics. Simulation results on BA trees (red, green and blue lines) with average degree $\left<k\right>=2$ and of size $N_0=10^4$ nodes and on BA networks (leftmost, brown line) of size $N_0=10^4$ nodes and with average degree $\left<k\right>=4$. The initially infected node was the largest hub in all cases and for each run a new tree or network was generated. The colors blue and green correspond to the average of  $10^4$ (blue line) and then further $4\cdot 10^4$ ($5 \cdot 10^4$ together) runs (green line) on BA trees for $\alpha=2$, whereas the red curve to the average of $5\cdot 10^4$ runs on BA trees for $\alpha=2.5$, and the brown curve is the average of $10^4$ runs on BA networks for $\alpha=2$. }}
\label{ugrasalphaketto}
\end{figure}

BA networks are locally tree-like objects \cite{Dorogovtsev}, as they have small number of loops, which are usually long \cite{loops}. However, even a small number of loops has a major impact on the spreading, as we show it for the $N_{\bar{t}}$ averages in Fig. \ref{ugrasalphaketto}. For stationary power-law dynamics with $1 <\alpha \leq 2$ on finite systems, the $N_{\bar{t}}$ average curves can be very different for BA trees and BA networks with loops (in  Fig. \ref{powwtossz}).

The blue and green  $N_{\bar{t}}$ spreading curves in Fig. \ref{ugrasalphaketto} are not smooth, jumps can be seen in the averages. These two curves are associated with stationary spreading on BA trees with exponent $\alpha=2$ and are the average of 10000 and further 40000 runs. Comparing the curves of the first 10000 (blue) and the first 50000 runs (green) for $\alpha=2$, one can observe that although old jumps start to shrink, new ones emerge. It is then expected that by increasing the number of runs, the jumps do not vanish for BA trees and for $\alpha \leq 2$, however, the red curve (BA tree, $\alpha=2.5$) and the brown curve (BA network, $\alpha=2$) in Fig. \ref{ugrasalphaketto}  and all the curves of BA networks even for $\alpha<2$  in Fig. \ref{powwtossz} are smooth in the interesting regime, i.e., except of the very late stage of the process.

\begin{figure}[H]
\centering
\includegraphics[width=8cm]{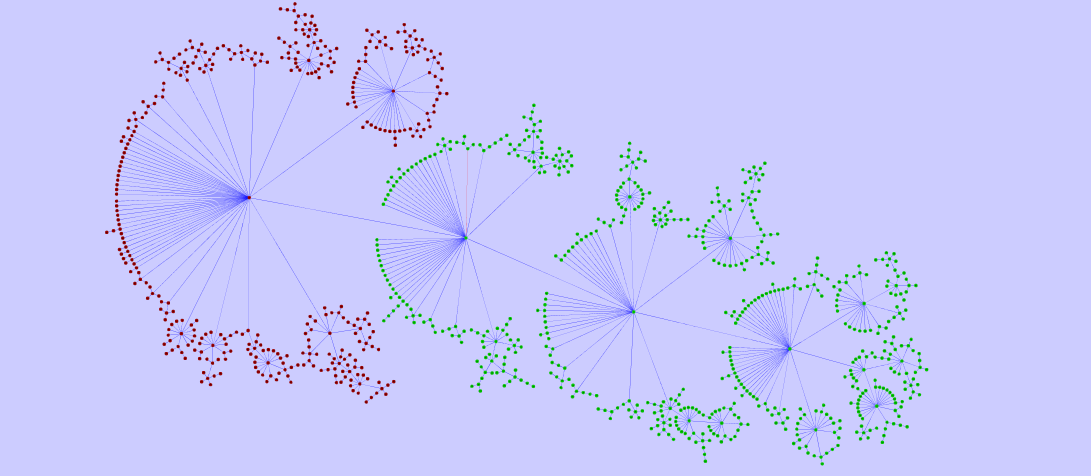}
\caption{\textit{Illustration of a BA tree of size 1000 nodes and the bottleneck effect. The red nodes are infected very fast, however, the only edge between the set of infected and susceptible (green) nodes can be activated at a very late time causing a jump in the average curve.}}
\label{bottleneck}
\end{figure}

By analysing the simulation runs the bottleneck effect turns out to be the cause of this phenomenon, which is illustrated by Fig. \ref{bottleneck}. When a part of the network gets entirely infected, and there is only one edge in the aggregated network between the set of infected and the susceptible nodes, the waiting time for the next infection can be extremely long. The reason is that the waiting time can have no expectation value and due to the lack of loops, the bottleneck cannot be circumvented.
Indeed, for finite trees, the existence of the mean of the WTD guarantees the well defined average spreading curve, whereas if no mean exits for the waiting time, due to the reasons mentioned above, i.e., the lack of loops meaning that the spreading has to pass each bottleneck, the curve cannot be smoothened by increasing the averaged simulation runs.
For power-law stationary dynamics, the WTD for $t>t_{min}$ is proportional to $\frac{1}{t^{\alpha}}$, thus for $\alpha \leq 2$ the WTD has no finite expected value, but for $\alpha >2 $ it has. 
Hence, for $\alpha <2$ the jumps do not vanish by increasing the number of runs, that is, actually no average exists. (For a more precise argumentation see \emph{Appendix B}.) Similar effect occurs on the Cayley tree too. 

The $\bar{N}_t$ average is well defined for $\alpha >1 $ as presented in \emph{Appendix B}, while for $\alpha \le 2 $, the non-existence of the $N_{\bar{t}}$ average, which is often the quantity of interest, makes it difficult to describe the spreading this way.  On the other hand the jumps are absent in the $N_{\bar{t}}$ averages of BA networks with loops (except the very late stage of the process), if the initially infected node is the hub or a node with larger degree. For finite trees the lack of meaningful averages is independent of the degree of the initially infected node if $\alpha < 2$, showing that the effect is merely due to the absence of loops. This clearly demonstrates the difference between trees and BA networks if the temporal processes are bursty. Despite BA networks are locally tree-like, the existence of  loops or the lack of bottlenecks seems to be crucial to having meaningful $N_{\bar{t}}$ averages if the IETD of the bursty processes is strongly fat-tailed. Thus the emergence of jumps in the averages can also be regarded as a nice interplay of temporal and topological effects. 

There is a subtlety regarding the infection of low degree nodes in BA networks with loops. Let us first consider the case when the spreading starts from such a node with degree $k_0$. For power-law, stationary dynamics (see \emph{Appendix C} for details) the average first infection time is finite if $\alpha> 1+\frac{1}{k_0}$ (Eqn. \eqref{C 6}), meaning that for $\alpha$ below this threshold, there is a jump in the spreading curve for any averaging. Starting the process from a large degree node as done in the above simulations the alternative routes enabled by the loops do not allow for the emergence of a bottleneck up to the very late stage, when only some of the low degree nodes remain susceptible. For example, it can be shown by similar reasoning to that in \emph{Appendix C}, the average infection time of the very last node diverges for sufficiently small $\alpha >1$. However, the steps induced by the infection of such low degree nodes appear only in the very late stage of the process thus they do not influence the interesting part of the spreading curve.

\section{Non-stationary dynamics on BA networks}

We have already emphasized that the use of \eqref{wtd} is allowed if the temporal processes are assumed to have been existing for infinitely long time. In many networks several effects (e.g. daily periodicity in  communication networks)  hinder the system to reach stationarity, therefore it is essential to study the effect of non-stationary dynamics as well, which, however, has not gained much attention in the literature so far. Non-stationary processes can be realized in many ways, the one we choose is simple and of practical importance. We define the temporal network as follows. 
For each link in the aggregated network, we define a starting time, when the first contact is established between the nodes in the temporal network corresponding to the link. From this time the contacts follow each other according to the IETD. The starting time has different values for the edges as it is equal to $-T_0- \zeta$, where $\zeta$ is a uniformly distributed random uncorrelated variable between  $[-\frac{\tau}{2}, \frac{\tau}{2}]$ and  $T_0>\frac{\tau}{2}$  is a non-negative number. As the first infection happens at $t=0$,  $T_0$ is the average age of the processes at the initial infection.
If the SI infection is simulated, similarly to the way presented in \emph{Section 4} we can deal with events only between susceptible and infected nodes. The only difference is, that if node $A$ is infected at time $t$, the possible infection times of its neighbors are generated by adding random numbers corresponding to the IETD from $-T_0- \zeta$ until the sum is larger than $t$. 
If the temporal process is Poissonian, there is no difference in the stationary and the non-stationary dynamics defined above. If the IETD is a modified exponential, the difference between the stationary and non-stationary behaviour should vanish rapidly with increasing $T_0$, as these distributions have fast convergence to zero for large times. These expectations are supported by numerical simulations, even for $T_0/\tau =1$ the stationary and non-stationary curves are notably close to each other.

\begin{figure}[H]
\centering
\includegraphics[width=7.5 cm]{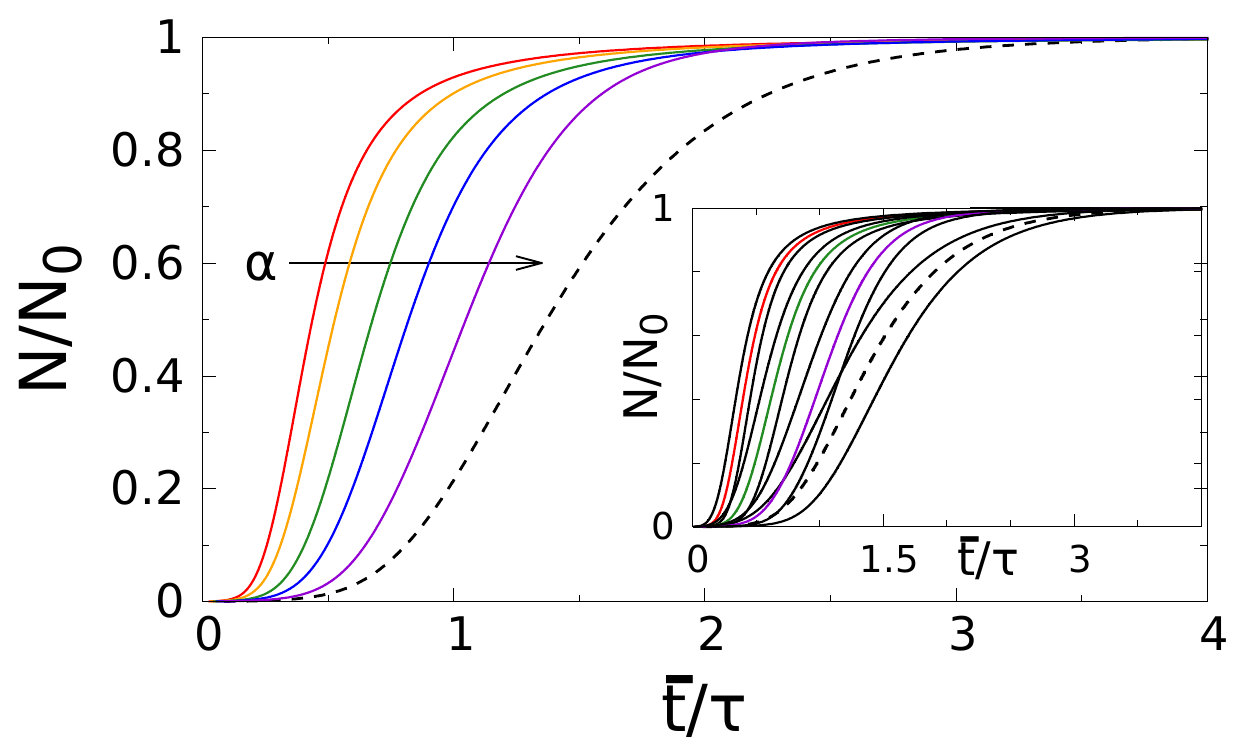}

\tiny
\caption{\textit{ Fraction of infected nodes $N/N_0$ vs average time $\bar{t}$ (measured in units of mean inter-event time $\tau$) spreading curves of pure exponential  (black, dashed line)  and Pareto (continuous, colored lines)  IETD governed non-stationary dynamics of age  $T_0=1$. Simulation results on BA networks of size $10^4$ nodes and with average degree $\left<k\right>=4$. The initially infected node was chosen at random in all runs in a way that its degree $k_0$ was larger than 25.  Averages of $2.5 \cdot10^3$ runs in each case and for every fifth run a new network was generated. The colors red, orange, green, blue and violet correspond to $\alpha=1.111, 1.154, 1.25, 1.4 \text{ and } 2$ and $t_{min}=0.1, 0.133, 0.2, 0.286 \text{ and } 0.5$  (in units of $\tau$)  lower cut-offs respectively. The arrow shows increasing $\alpha$ values with the exception of the dashed curve (Poisson process), for which $\alpha=1$. The inset indicates the fluctuations of the curves associated with $\alpha=1.111, 1.25 \text{ and } 2$ with continuous black lines.}}
\label{powietossz}
\end{figure}

Non-stationarity has a remarkable effect on the spreading for power-law IETD governed dynamics. This is reflected both in the spreading speed and in the convergence to the asymptotic behaviour ($T_0 \to \infty$). The response to tuning $\alpha$ is the opposite to what we saw in the stationary case for small $T_0$, i.e., decreasing the power-law exponent the infection accelerates Fig.  \ref{powietossz}. Consequently, there has to be a crossover from acceleration from deceleration as a function of $T_0$. 

\begin{figure}[H]
\centering
\includegraphics[width=6.25 cm]{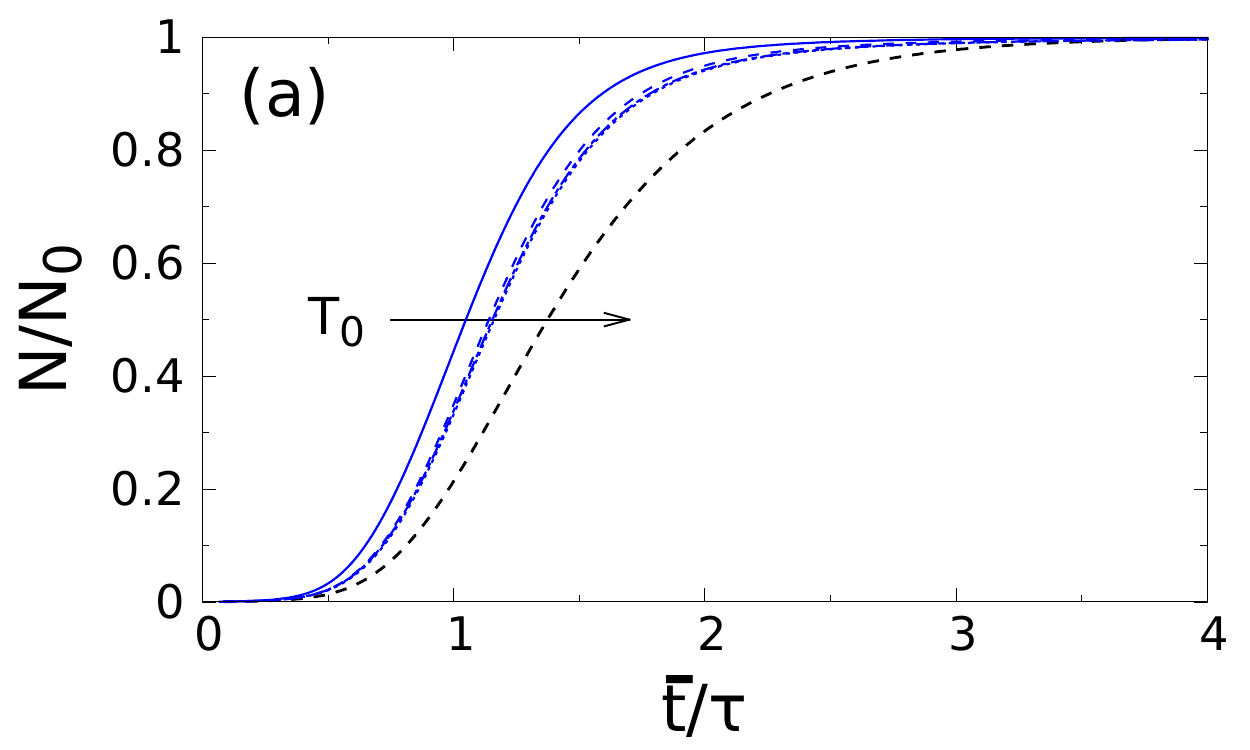} \includegraphics[width=6.25 cm]{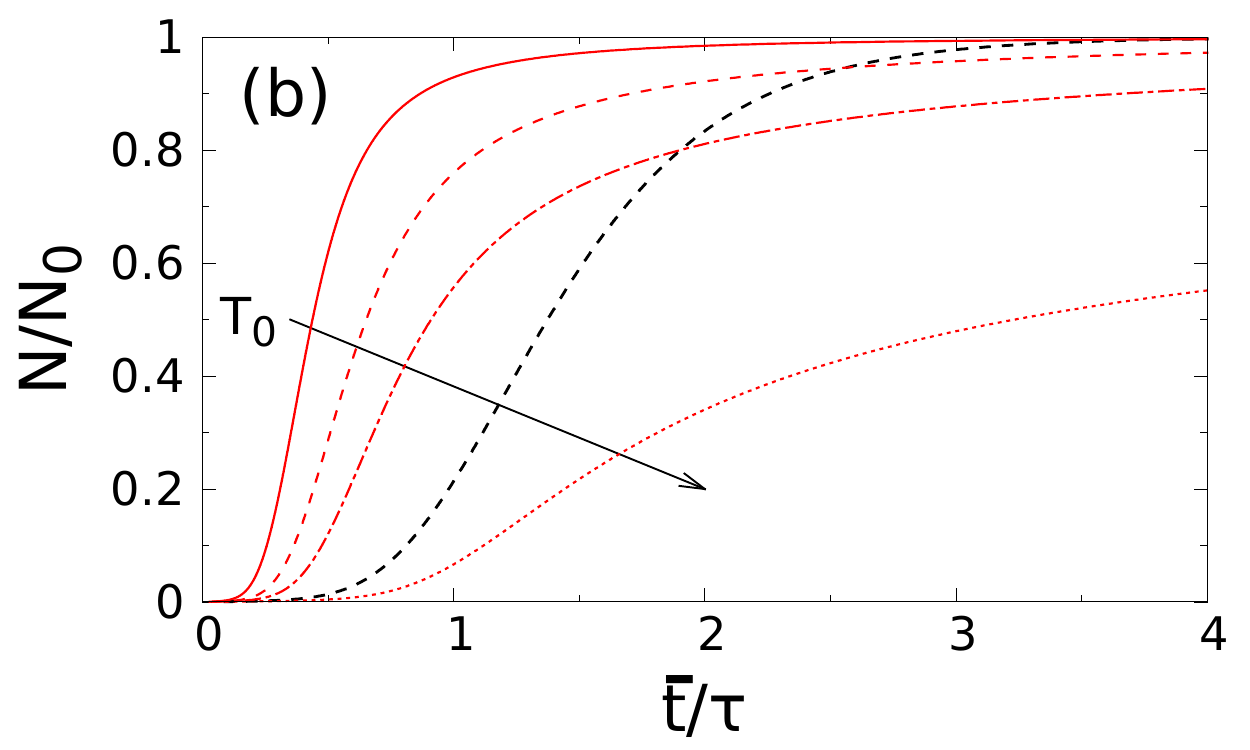}
\caption{\textit{{ Fraction of infected nodes $N/N_0$ vs average time $\bar{t}$  (measured in units of mean inter-event time $\tau$) curves of pure exponential  (black, dashed line)  and Pareto (colored lines)  IETD governed stationary ($T_0= \infty$}) and non-stationary dynamics with some given values of $T_0$. ($T_0=1, 10, 100$.) Simulation results on BA networks of size $N_0=10^4$ nodes with average degree $\left<k\right>=4$. The initially infected node was chosen at random in all runs in a way that its degree $k_0$ was larger than 25.  Averages of $2.5 \cdot10^3$ runs in each case. The colors red (a) and blue (b) correspond to $\alpha=1.111 \text{ and } 2$ exponents and $t_{min}= 0.1 \text{ and } 0.5$  (time in units of $\tau$) lower cut-offs. The curves for $T_0=1, 10, 100 \text{ and } \infty$ are drawn with continuous, dashed, dotted-dashed and dotted lines respectively and the arrows show increasing values of $T_0$ with the exception of black dashed curve, for which $T_0= \infty$.}}
\label{powietossztobb}
\end{figure}

In order to understand this crossover, we have to study the convergence of the spreading curves to the stationary averages.
According to  Fig.  \ref{powietossztobb}, the smaller the $\alpha$ exponent is, the more time is needed to reach stationarity. For $T_0=100$ the blue curve ($\alpha=2$) concurs with its  stationary position, whereas the red curve ($\alpha=1.111$) is far away from it.

\begin{figure}[H]
\centering
\includegraphics[width=6.5 cm]{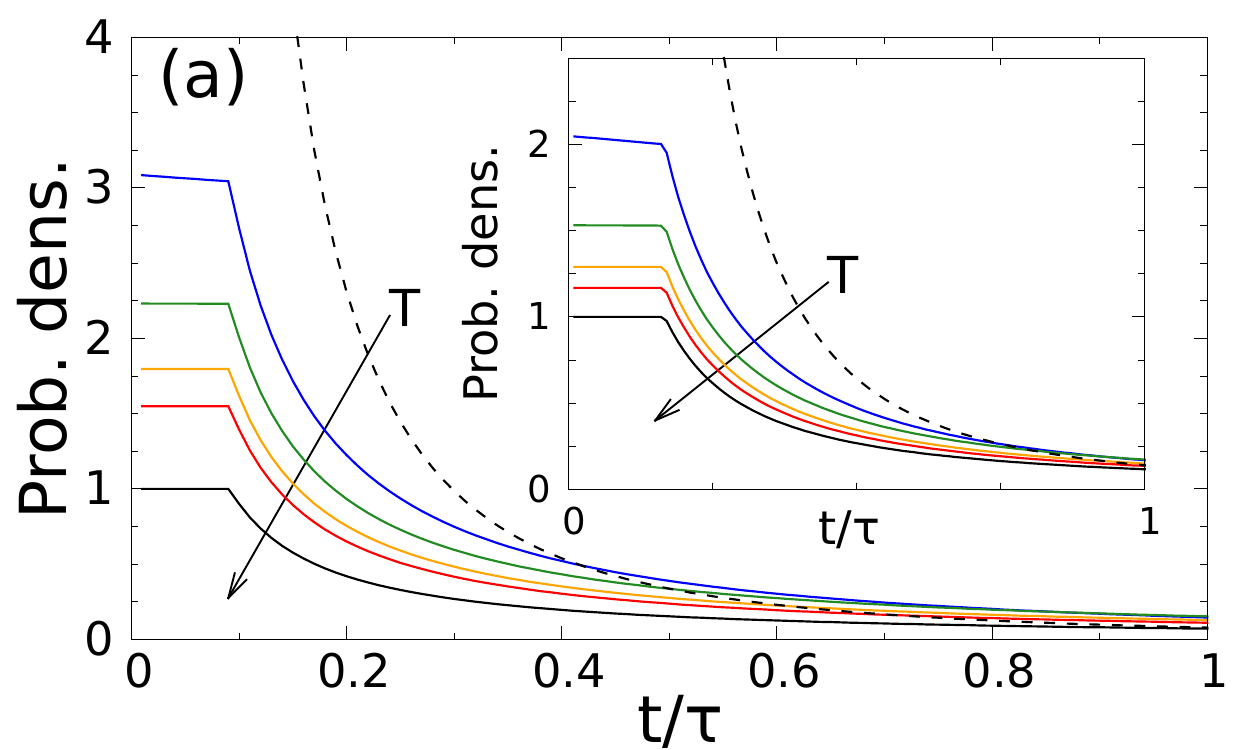} \includegraphics[width=6.5 cm]{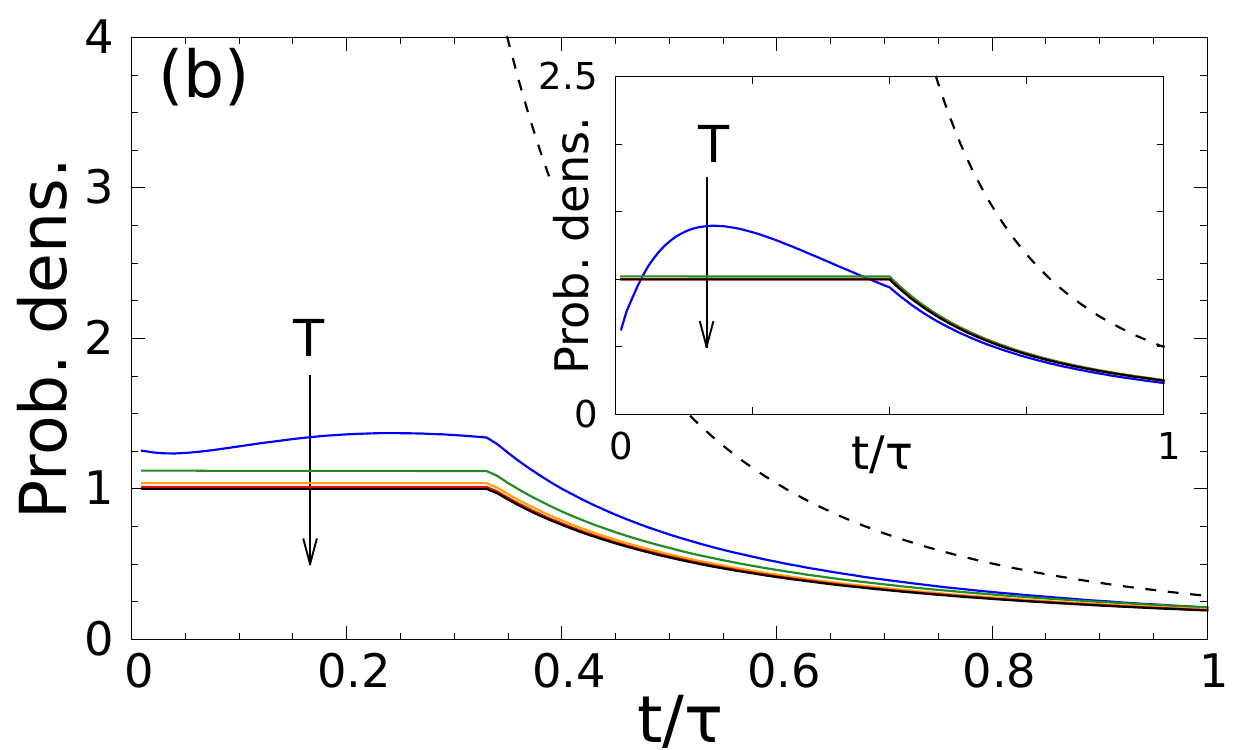} 
\tiny
\caption{\textit{Age dependent waiting time distributions for processes of different ages $T$ and with Pareto inter-event time distributions with different values of the $\alpha$ exponent. The arrows show increasing values of $T$; the blue, green, orange and red lines correspond to ages  1, 10, 100, 1000 in units of $\tau$, and the black dashed and black continuous lines to the inter-event time distribution ($T=0$) and the stationary waiting time distribution ($T=\infty$). Distributions for $\alpha=1.1$ and $\alpha=1.2$ are displayed in the left figure (a) and in the left inset, and distributions for $\alpha=1.5$ and $\alpha=2$ in the right figure (b) and in the right inset respectively.}}
\label{distkonv}
\end{figure}

To understand the (occasionally) slow convergence of the spreading curves and the slow-fast transition for small $\alpha$-s, we first study how the age dependent WTD converges to its limit distribution. For this, let us consider an arbitrary sequence of events so that the inter-event times between them are iid random variables corresponding to the Pareto distribution, and let the first event happen at $t=0$. $H(T, \xi )$ denotes the probability that at time T, for the next event at most  $\xi$ has to be waited, hence for $T \to \infty$ H tends to the cumulative distribution of the stationary WTD.
In theory, $H(T, \xi)$ can be calculated by the following formula \cite{Feller}:
\begin{equation}\label{H}
H(T, \xi)=\int_0^T U\{ \,dx \} \left( F(T-x+\xi)-F(T-x) \right),
\end{equation}
where $F$ denotes the cumulative IETD and $U(t)$ is the average number of events in $[0, t]$ closed interval. Note that $U$ has an atom of unit at the origin, hence the measure of the $[0, 0+\varepsilon]$ interval is 1, if $\varepsilon \to 0^+$.
For $t \to \infty$,  $\frac{U(t)}{t}$ equals $\tau^{-1}$ but for finite times $U(t)$ is not necessarily equal to $1+\frac{t}{\tau}$. For any $t>0$, $U$ can be calculated as \cite{Feller}

\begin{equation}\label{U}
U(t)=1+\int_0^t  U(t) F(t-x)\,dx ,
\end{equation}
where $F$ is the cumulative IETD again.
The computation of $U(t)$ and hence $H(T, \xi)$  is possible only numerically if the IETD-s are Pareto distributions.  In Fig. \ref{distkonv} the age dependent waiting time distributions are displayed for different $T$ and $\alpha$ parameters together with the IETD and the time independent WTD that correspond to $T=0$ and $T=\infty$, respectively.

From Fig. \ref{distkonv} one can see that indeed the smaller the $\alpha $ is, the slower convergence is resulted. Another interesting aspect is, however, that for small $\alpha$ and small $T$, the probability density is relatively large for small $\xi$.  It is natural to assume that this density accounts for the fast spreading seen in Fig. \ref{powietossz} in the case of young processes. For small $\alpha$ the IETD for times not much larger than the lower cut-off also has large density, which is shifted in the interval $[0, t_{min}]$ and only slowly redistributed in other parts of the $(t_{min}, \infty)$ interval as the age of the process $T$ is increased. If the process is old enough the WTD for small times is suppressed and the tail has a stronger impact on the overall spreading.


\newpage
\section{Summary and Discussion}

In this paper we applied the SI model to study spreading phenomena associated with bursty processes on networks and observed rich variety of behaviour. For stationary processes we found that bursty, power-law inter-event time distributions can lead to cardinally slow spreading for power-law exponents below a threshold value. This observation is based on analytic calculations with Cayley trees and numerical simulations using BA networks. The slowing down is in contrast with \cite{egzakt}, however, two remarks should be made here. First, the underlying network in \cite{egzakt} was basically a complete graph, while we considered here more realistic sparse graphs. Second, the clocks of the process were put there onto the nodes, while we had them on the links. It should also be mentioned that although we see always a slowing down with Pareto IETD as compared to the case with the exponential IETD with the same lower cutoff, there is a crossover from slowing down to acceleration if we compare with the Poissonian case (see Fig. \ref{Cfanagyido}).

We pointed out an interesting difference between finite real tree graphs and locally tree-like networks from the point of view of spreading. The existence of rare loops has turned out to be crucial to avoid hectic jumps in the $N_{\bar{t}}$ spreading curves if the waiting time distribution has no finite mean, therefore only locally tree-like networks and trees can behave in a very different manner under bursty processes, which also implies that tricky methods approximating sparse networks with few loops by trees are not always allowed to use.

The role of non-stationary in spreading phenomena has gained some attention only recently. In the work of Holme et al. \cite{HL} the beginning and end times of dyadic communication sequences  proved to be an important element in the spreading speed and in the paper of Rocha et al. \cite{SIR} non-stationary contact processes resulted in faster power-law spreading for short times in SI models than the Poisson process. 
In our paper we carried out a systematic investigation of non-stationary processes, which illuminates many important aspects of the question of non-stationarity. We showed that power-law governed, non-stationary processes of young age can cause very rapid spreading in agreement with \cite{SIR} even for power-law exponents that would result in slow spreading in the stationary state, hence the age of the processes has a strong influence on the outcome of the spreading if the inter-event time distribution is strongly heavy-tailed. Increasing the age of the processes the convergence of the non-stationary spreading curve to the stationary one is very slow for small enough power-law exponents, which is the result of the slow convergence of the age dependent waiting time distribution. By numerically calculating these distributions, we managed to show that for small power-law exponents and small waiting times the probability density can be large for young processes. As for bursty dynamics the time scale of the convergence can be much larger than the time scale of a typical inter-event time, the age of the temporal network at the start of the spreading is expected to play an important role in the spreading process. 

We have demonstrated that the question of the speed of spreading is a complex one. The IETD, the topology and the age of the temporal process have all impact on it. As for the IETD, it is clear that the early time is often more important than the tail. This is understandable, as if several possibilities are for finding a spreading path, it is the one with the shortest waiting time, which is selected. However, sometimes the tail becomes dominant. This is the case, if a bottleneck is formed as on finite trees. This already demonstrates the importance of the topology. Last but not least the non-stationarity of the processes is crucial, as depending on the age of the process the same IETD can lead to an accelerated or decelerated spreading. We think that the rather complex picture emerging due to the above factors may be the source of the controversial observations, as all these factors are present under empirical conditions.

\ack
This work was prepared within the framework of the project FuturICT.hu (grant no.: T\'{A}MOP-4.2.2.C-11/1/KONV-2012-0013) and it was partially supported by DATASIM FP7 27833 project and the FiDiPro program by TEKES 125587 of Finland.

\appendix
\section*{Appendix A-\emph{The $N_{\bar{t}}$ and the $\bar{N}_t$ averages}}

In our paper we distinguished the two basic ways of calculating averages and we have seen that the $N_{\bar{t}}$ average can be meaningless under some circumstances, whereas the $\bar{N}_t$ average is usually well-behaved.
These two averages, however,  are not independent from each other as the probabilities $\mathds{P}\{N_t < N \}$ and $\mathds{P} \{t<t_N \}$ have to be equal, that is,
\begin{equation}\label{nttn}
\mathds{P}\{N_t < N \}=\mathds{P} \{t<t_N \} \text{ or } \mathds{P}\{N_t < N \}=1-\mathds{P} \{t_N<t \}, \tag{A.1}
\end{equation}
where both $N_t$ and $t_N$ are stochastic processes; $N_t$ for a given $t$ is the number of infected nodes until time $t$ and $t_N$ for a given $N$ is the time elapsed until the $N$-th infection.
For finite networks any moments of the $\mathds{P} \{N_t<N\}$ distribution are finite for any time $t$ and even in infinite systems the previous probability should converge to one exponentially for large $N$ for any time. If bursty processes are considered the growth in the number of infected nodes can be very slow, i.e., for a fixed $N$ the difference between the probabilities $\mathds{P}\{N_t < N \}$ and $\mathds{P}\{N_{t+t'} < N \}$ can be very small even for large $t'$. As \eqref{nttn} holds, it follows that the distribution $\mathds{P} \{t_N<t \}$ can be very broad resulting in divergent moments as seen in connection with spreading on trees in \emph{Section 4}.
For the existence of the $N_{\bar{t}}$ average the following criterion can be derived:
According to Eqn. \eqref{nttn}  $\frac{d}{dt} \mathds{P}\{N_t < N \}=-\frac{d}{dt}  \mathds{P} \{t_N<t \}$, hence
\begin{equation}\label{fnt}
\bar{t}_N=\int_0^{\infty} t \cdot \frac{d}{dt}  \mathds{P} \{t_N<t \} \, dt= - \lim_{t \to \infty} t \cdot \mathds{P}\{N_t < N \}+ \int_0^{\infty}\mathds{P}\{N_t < N \} \, dt, \tag{A.2}
\end{equation}
which is finite if 
\begin{equation}\label{limesz}
\lim_{t \to \infty} t \cdot \mathds{P}\{N_t < N \} = 0, \tag{A.3}
\end{equation}
that is for a fixed $N$ and for large $t$, $\mathds{P}\{N_t < N \}$ asymptotically converges faster to zero than $1/t$.

\appendix
\section*{Appendix B-\emph{Spreading on trees}}

 By a simple calculation we now address the question of whether the jumps seen in the $N_{\bar{t}}$ curves of trees in \emph{Section 4} can be smoothened by either tuning $\alpha$ or increasing the number of runs.

Consider a BA tree of nodes $M$, where $M$ is finite. For a visible jump after $N$ runs it is necessary that there exist at least one run in which the infection on the link connecting the two sides of the bottleneck happens at a larger time than $a N$  (time measured in units of $\tau$ where $a > 1$) as the typical time scale of the spreading process is determined by $\tau$.  It is also necessary that all the nodes in one side of the bottleneck be either infected with infection times much less than some constant times $\tau$ or be infected only later than the node in the bottleneck. To calculate the probability of such a jumpy run let us  assume that all the nodes in the infected part of the network are direct neighbors of the initially infected node. It is not difficult to see, that the argumentation remains valid without this assumption. Let $p$ denote the probability that a node gets infected at a larger time than $aN$ and let $q$ be the probability that it gets infected before 1 (in units of $\tau$). Due to the additional assumption $p$ can be derived from the WTD and $p=\frac{k}{(aN)^{\alpha-1}}$, where $k$ is an $\alpha$ dependent constant.
The probability of a jumpy run is 
\begin{equation}\label{jumpy}
\begin{aligned}
\mathds{P}\{ \text{jump in the av. from 1 run} \}&=p\left( q^{M-1}+ \binom{M-1}{1}p q^{M-2}+\binom{M-1}{2}p^2 q^{M-3} \cdots p^{M-1} \right) \\
&=p (p+q)^{M-1} ,
\end{aligned}
\tag{B.1}
\end{equation}
where the binomial terms correspond to being more than one long waiting periods and $M$ could be replaced by the number of nodes in one side of the bottleneck, which is at most $M$.
Then the absence of jumps in the average of $N$ independent runs is
\begin{equation}\label{jumpy2}
\mathds{P}\{ \text{No jumps after  N runs} \}= (1-p (p+q)^{M-1})^N=  \left( 1-\frac{k}{(aN)^{\alpha-1}} (\frac{k}{(aN)^{\alpha-1}}+q)^{M-1} \right)^N. \tag{B.2}
\end{equation}
It is easy to see that the expression in Eqn. \eqref{jumpy2} has the same limit as Eqn. \eqref{jumpy3} as $M$ is finite.
\begin{equation}\label{jumpy3}
 \lim_{N \to \infty} \left(1-\frac{k'}{(aN)^{\alpha-1}}\right)^N. \tag{B.3}
\end{equation}
This limit is 0 if $\alpha < 2$, and is 1 for $\alpha >2$, that is smooth spreading curves are expected if $\alpha >2$ as seen also in the paper in \emph{Section 4}, meaning that
 for $\alpha <2$ the jumps do not vanish by increasing the number of runs, that is, actually no average exists.

\begin{figure}[H]
\centering
\includegraphics[width=6.5 cm]{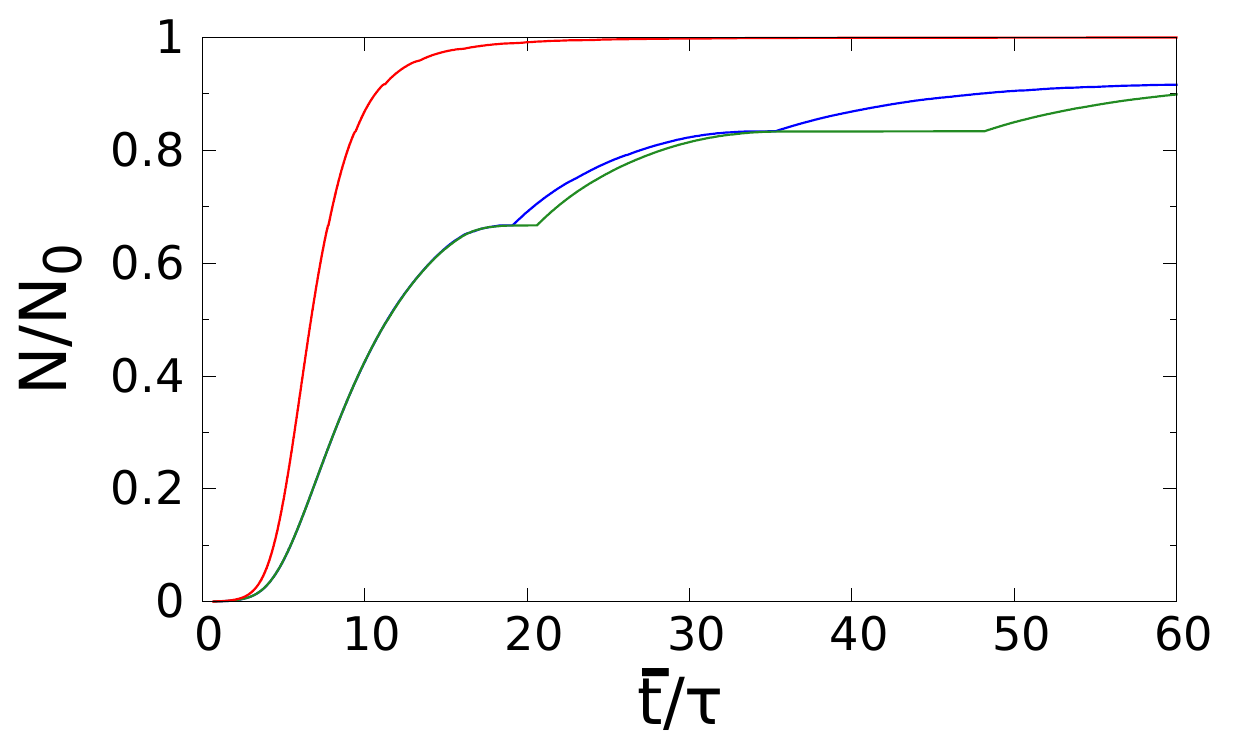} 
\tiny
\makeatletter 
\renewcommand{\thefigure}{B\@arabic\c@figure}
\makeatother
\caption{\textit{Fraction of infected nodes $N/N_0$ vs average time $\bar{t}$ (measured in units of mean inter-event time $\tau$) spreading curves of power-law governed stationary dynamics on a Cayley tree of size $N_0=12286$ and with $k=3$. Averages of $10^4$ (middle, blue line) and then further $4\cdot 10^4$ ($5 \cdot 10^4$ together) runs (rightmost, green line) for $\alpha=2$ and average of $5 \cdot 10^4$ runs for $\alpha=3$  (leftmost, red line). For each run  the initially infected node was the central node. }}
\label{ugrasCfa}
\end{figure}

\begin{figure}[H]
\centering
\includegraphics[width=7 cm]{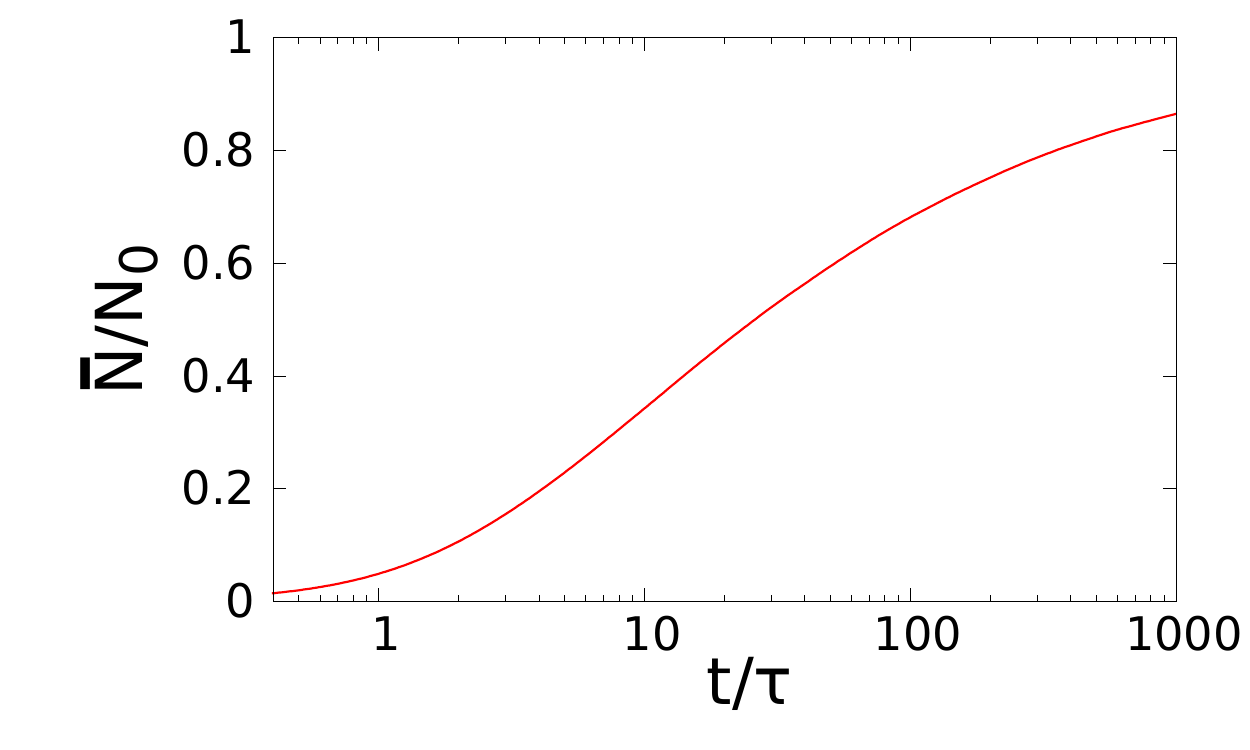} 
\tiny
\makeatletter 
\renewcommand{\thefigure}{B\@arabic\c@figure}
\makeatother
\caption{\textit{Average fraction of infected nodes $\bar{N}/N_0$ vs time $t$  (measured in units of mean inter-event time $\tau$) spreading curve of power-law governed stationary dynamics with $\alpha=1.4$ on a BA trees with average degree $\left<k\right>=2$ and of size $N_0=10^4$ nodes. Average of $5 \cdot 10^3$ runs. For each run the initially infected node was the largest hub. To calculate the average logarithmic binning was applied with 100 bins for each order of magnitude.}}
\label{BAfaVA}
\end{figure}

With the following figures we wish to confirm some of our statements in connection with the jumpy spreading curves. Although our calculation regarding the disappearance of the jumps has to hold for any finite trees, some confirmation based on simulations is also welcomed. Fig. \ref{ugrasCfa} further underpins the validity of this calculations presenting the average of simulation runs if the aggregated network is a Cayley tree and the $N_{\bar{t}}$ average is considered. According to this figure the jumps are absent if $\alpha >2$ but the are present if $\alpha=2$ or smaller. One can even observe that the jumps seem to happen always around the same level of infection, namely at  66.66\% and 82.5 \% which corresponds to evolving a bottleneck at either one of the neighbors of the initially infected node or at one of its second neighbors, as $k=3$.
We have already emphasized that for observing the jumps the $N_{\bar{t}}$ averages have to be considered. The $\bar{N}_t$ averages are always well-behaved since they represent existing averages as supported by Fig. \ref{BAfaVA}, which shows a spreading curve of a power-law governed  stationary dynamics with $\alpha=1.4 $ on BA tree. Clearly, the $N_{\bar{t}}$ average does not exist for this exponent, the $\bar{N}_t$ curve in Fig. \ref{BAfaVA} is, however, apparently smooth.

\appendix
\section*{Appendix C-\emph{First infection times}}

Here we derive a formula by means of which the average time of the first  infection can be calculated, if the initially infected node has $n$ neighbors and stationary, power-law dynamics is considered. From the mathematical point of view, we are interested in merely the mean of the minimum of $n$ iid ($Y_1, Y_2, Y_3...Y_n$)  random variables, denoted by $X_n=min(Y_1, Y_2, Y_3...Y_n)$.
For $n=2$,  $ E(X_2) $ is defined by the following integrals in which $p(y)$  is the density function of $Y$: 
\begin{equation}\label{C 1}
E(X_2)=2\int_0^{\infty}\int_x^{\infty} p(y) \,dy \text{ } x p(x) \,dx. \tag{C.1}
\end{equation}
For $n=3$ the domain of integration is easy to see and to generalize, yielding $\forall n >1$ :
\begin{equation}\label{C 2}
E(X_n)=n\int_0^{\infty} \left(\int_x^{\infty} p(y) \,dy \right)^{n-1} x p(x) \,dx. \tag{C.2}
\end{equation}

Considering stationary, power-law spreading, for the calculation of the first infection time  $ p(y)$ in the formulae above has to be equal to $p'_{pow}(t)$, which is the WTD derived from the Pareto IETD.
\begin{equation}\label{C 3}
p'_{pow}(t)=
\begin{cases}
\frac{1}{\tau},                                                                   & \text{ if }\ 0 \leq t \leq t_{min}\\
\frac{1}{\tau} t_{min}^{\alpha}\frac{1}{t^{\alpha}},     & \text{ if }\ t > t_{min}.
\end{cases} \tag{C.3}
\end{equation}
Hence
\begin{equation}\label{C 4}
\int_x^{\infty} p'_{pow}(y) \,dy=
\begin{cases}
\frac{1}{\alpha-1} \frac{1}{\tau} \frac{t_{min}^{\alpha}}{x^{\alpha-1}},          & \text{ if }\ x \geq t_{min}\\
\frac{1}{\alpha-1} \frac{t_{min}}{\tau}+\frac{1}{\tau}\left( t_{min}-x \right),   & \text{ if }\ 0 \leq x \leq t_{min}.\\ 
\end{cases} \tag{C.4}
\end{equation}
Then, by \eqref{C 2} one  obtains:
\begin{equation}\label{C 5}
E(X_n)=n\left[  \int_0^{t_{min}}\!\!\! \left( \frac{1}{\alpha-1} \frac{t_{min}}{\tau}+ \frac{\left( t_{min}-x \right)}{\tau}  \right)^{n-1}\!\! \frac{x}{\tau} \,dx + \int_{t_{min}}^{\infty} \left(\frac{1}{\alpha-1} \frac{1}{\tau}\frac{t_{min}^{\alpha}}{x^{\alpha-1}}\right)^{n-1}\!\!\frac{x}{\tau} \frac{t_{min}^{\alpha}}{x^{\alpha}} \,dx \right]. \tag{C.5}
\end{equation}
It is easy to see that the condition for the existence of expected value is expressed as:
\begin{equation} \label{C 6}
 \alpha > \frac{1}{n}+1. \tag{C.6}
\end{equation}

Omitting tedious calculations, the average first infection time is expressed as follows:
\begin{equation}\label{C 7}
E(X_n)= \tau \left[ \frac{1}{(n+1)}\left[ 1-\left(1+(n+1)(\alpha-1) \right) \frac{1}{\alpha^{n+1}}\right] + \frac{n}{n \alpha-(n+1)} \frac{(\alpha-1)^2}{\alpha^{n+1}}  \right] \tag{C.7}
\end{equation}
It is worth taking a look at some limits of \eqref{C 7}.  If $n$  is sufficiently large, due to the smallness of the  $\frac{1}{\alpha^{n+1}}$ type terms, $E(X_n)$ can be approximated as
\begin{equation}\label{C 8}
E(X_n) \approx \frac{\tau}{n+1} \tag{C.8}
\end{equation}
According to ~\eqref{C 8} for large $n$ the first infection happens earlier on average in the  Pareto than in the Poissonian case, since in the latter $E(X_n)=\frac{\tau}{n}$. By evaluating \eqref{C 7} it can also be seen that for fixed $n$ the power-law average infection time is smaller than the Poissonian for small power-law exponents and larger for large exponents.

\end{document}